\documentclass[aps,twocolumn]{revtex4}  

\usepackage{graphicx}

\newcommand{\bfp}{}

\newcommand\beq{\begin{equation}}
\newcommand\eeq{\end{equation}}
\begin{document}

\title{Membrane Heating in Living Tissues \protect\\
 Exposed to Nonthermal Pulsed EM Fields }
\author{
V. Pierro, A. De Vita, R. P. Croce \protect\\
and I. M. Pinto,~
\thanks{V. Pierro, R. P. Croce and I. M. Pinto  are with the Department of Engineering, University of 
Sannio, C.so Garibaldi 107, 82100 Benevento, Italy. A. De Vita is with Centre for Research and Technological Innovation RAI- Radiotelevisione Italiana,
C.so E. Giambone 68, 10135 Torino,  Italy.
}}


\markboth{IEEE Transactions on Plasma Science,~Vol.~X, No.~Y, January~ZZZZ}
{Shell \MakeLowercase{\textit{et al.}}: Membrane Heating in Living Tissues...}

\begin{abstract}
\boldmath
A bio tissue model consisting of multilayer spherical cells including
four nested radial domains (nucleus, nuclear membrane, cytoplasm and plasma membrane) 
is worked out to derive the cell heating dynamics in presence of membrane
capacitance dispersion under pulsed electromagnetic exposure. 
Two possible cases of frequency-dependent membrana models are discussed:
plasma and nuclear membranes are dispersive, 
only the nuclear memebrane is dispersive . In both models an high 
localized  heating of the membranes occurs,  
without significant temperature rise in the cytoplasm and nucleoplasm.
\end{abstract}
\maketitle


\section{Introduction}
%
Substantial efforts on behalf of the Scientific Community have been recently addressed
to the experimental assessment and theoretical investigation \cite{Garner} of the response of living
cells to ultrashort (sub-nanosecond) intense (MV/m) electromagnetic field pulses (see also \cite{tutorial} for a tutorial).
Cell exposure to pulsed EM fields in the msec-$\mu$sec range is known to produce
transient or permanent permeabilization (electroporation \cite{plasma_perm}) of the cytoplasmic
membrane, depending on pulse amplitude, through a peculiar breakdown phenomenon
occurring when the transmembrane potential difference exceeds some critical level ($\sim$1V).
Cell response to {\em ultrashort} pulses is markedly different. Basically, the integrity of the
cytoplasmic membrane is {\em not} directly affected, although the transmembrane potential
difference may largely exceed the poration breakdown threshold. 
Substantial permeabilization of the organelles, including the nucleus is observed instead, usually
triggering an apoptotic response (signaled, e.g., by externalization of phosphatidylserine
\cite{phosphat}), which eventually (typically several minutes after exposure) leads to membrane
dissolution and cell remnants removal by macrophages.
Such a mechanism may hold a potential for cancer treatment \cite{zap}. 
In fact, effective
selective destruction of several types of tumors, including skin melanomas and colorectal
carcinomas (two of the 'big killers'), up to complete remission, have
been reported after suitably tuned ultrashort EM pulse exposure \cite{melanoma}, \cite{other_cancers}.
This mechanism of action differs completely from that of microwave hyperthermia \cite{heating},
where tumor necrosis is induced by selective heating of the neoplastic tissue.
Exposure to ultrashort pulses, on the contrary, does {\em not} produce
any sensible macroscopic thermal response, nor (killing the cells through apoptosis) an
inflammatory response.
In order to understand the underlying physical mechanisms,
it is necessary to develop a model providing a good tradeoff between simplicity and realism.
Most studies are based on the cell model proposed by Schwan et al. 
\cite{Schwan_1}-\cite{Schwan_3} consisting of a single spherical cell (extracellular medium, 
plasma membrane and cytoplasm) whose electromagnetic constitutive parameters are treated as frequency independent.
On the other hand, it is well known that the membrane specific capacitance is {\em strongly} frequency-dependent 
\cite{Haydon}.
In \cite{DeVitaPS} we investigated the thermal response of a simple cell model (spherical homogeneous, with frequency dependent
membrane capacitance)  to a pulsed-electromagnetic field by solving the coupled
electromagnetic and heat-diffusion problems. 
Our findings suggest that whenever the pulse duration is small compared to the thermal
relaxation constant of the cell membrane, 
and the membrane capacitance drops to its low asymptotic high-frequency value  
in the pulse spectral bands, 
one may observe a steep increase in the membrane temperature, 
up to physiologically significant levels, 
the average (cytoplasm) temperature remaining essentially unaffected.
Similar conclusions were obtained  in \cite{Song} 
following a more sophisticated approach which combines  
Smoluchowski equation to describe membrane response, 
the heat equation and molecular dynamics simulations, 
to gauge the impact of localized membrane heating on membrane poration.\\
In this paper we extend the analysis of \cite{DeVitaPS} to the more realistic case
where the cell is part of a bio-tissue.
We also adopt a more general cell model
consisting of four radially nested domains (nucleus, nuclear membrane, cytoplasm and plasma membrane {\bfp supposed concentric}),
including frequency dependence of both the plasma and nuclear membranes.
{\bfp
In the paper the
membrane capacitive models used do not depend on voltage, this implies
that the analysis holds only when the external
electric fields are not strong or long enough to create membrane pores. 
}

In order to solve the pertinent electromagnetic boundary value problem, in the quasi-static limit,
we use Effective Medium Theory (henceforth EMT) throughout \cite{MixBook}. 
Numerical simulations based on this more realistic model confirm the occurrence of
a steep temperature raise in the plasma membrane, without any significant temperature
variation in the cytoplasm and nucleus. \\
The paper is accordingly organized as follows.
In Section II we use EMT to derive the effective
permittivity of a tissue; in Section 
III we derive the response of a cell inside a tissue, connecting
the local field to the external (impressed) field. In Section IV we solve 
the coupled electromagnetic and heat-diffusion problems for a 
(multilayer) spherical cell with linearized, dispersive 
nuclear and plasma membranes, embedded in a tissue 
and exposed to a (pulsed) monochromatic electromagnetic field, 
using a toy model for membrane capacitance dispersion.
Representative numerical results are briefly 
illustrated and discussed in Section V.
Conclusions follow in Section VI.
%
\section{Cells in a Tissue}
\label{dense_systems}
\begin{figure}[h]
\includegraphics[scale=1.0, width=9cm]{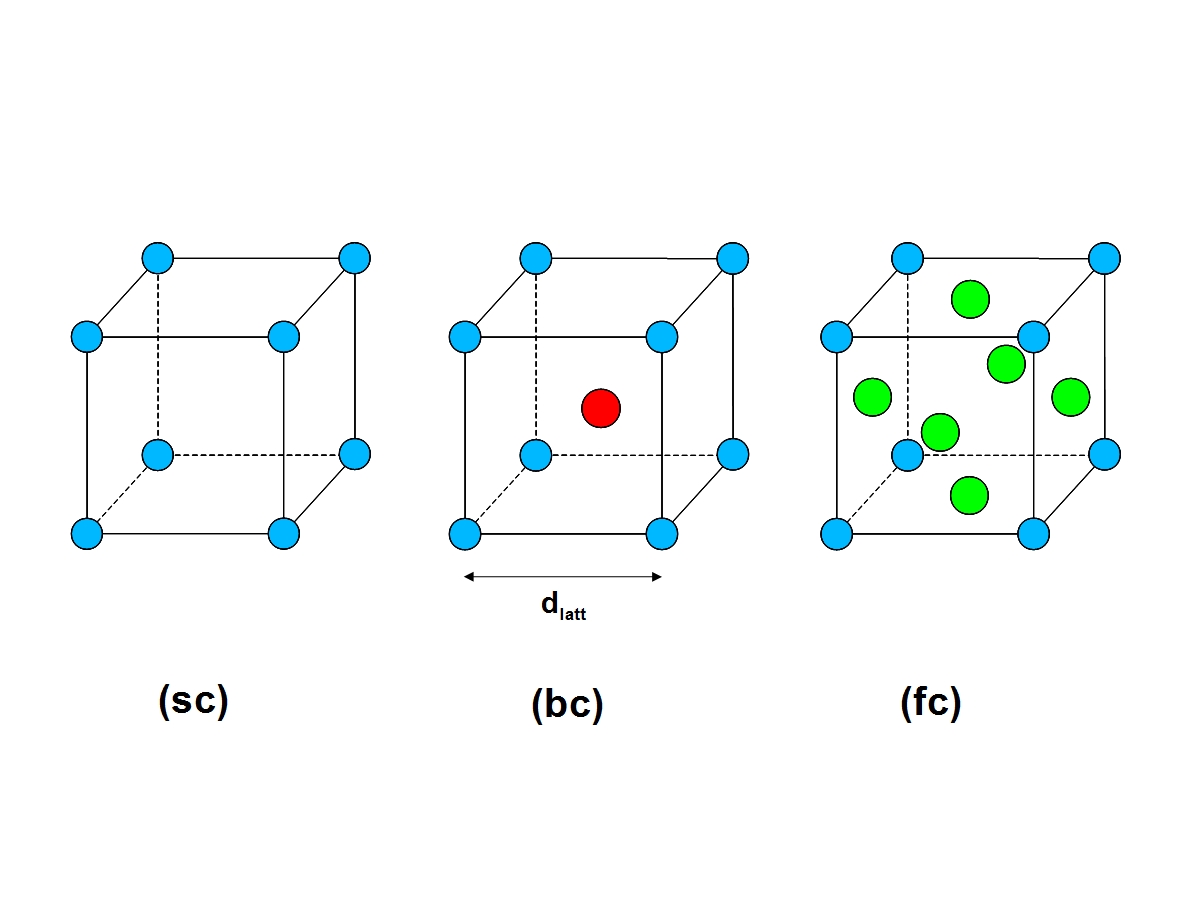}
\caption{
The simple-cubic (sc), body-centered cubic (bc), and face-centered cubic (fc) lattice models.
The lattice unit-cell side length $d_{latt}$ is displayed.
}
\label{fig1}
\end{figure}

%
Living tissues are regular assemblies  of densely packed cells. 
Cubic lattices, e.g., including the simple-cubic ({\em sc}), body-centered cubic ({\em bc}) and face-centered cubic ({\em fc}) 
arrangements sketched in Fig. \ref{fig1}, have been suggested as simple tissue morphology models \cite{tissue_cubic}.
%
%
The ratio between the cell radius $R_c$ and the lattice unit-cell side length 
$d_{latt}$ in Fig. \ref{fig1}  determines the cell volume fraction $f_c$,
\begin{equation}
f_c=\frac{4}{3}\pi~N_c\left(\frac{R_c}{d_{latt}}\right)^{3},
\label{eq:volume_frac}
\end{equation} 
$N_c$ being the number of bio-cells contained in the lattice unit-cell 
($1$, $2$ and $4$ for the  {\em sc}, {\em bc} and {\em fc} cubic lattices, respectively).
The maximum volume fractions correspond to {\em touching} cells, and
are $0.52$, $0.64$ and $0.74$ for {\em sc}, {\em bc} and {\em fc} cubic lattices, respectively.
{\bfp
In this paper we consider separate living cell (no touching cells) in the tissue.
The cells density reported above  
is the maximum volume fraction that can be obtained for a suitable value of ratio $R_c/d_{latt}$, as can be seen by equation (\ref{eq:volume_frac}).

In order to model dense tissue we approximate the cells arrangement like a face centered cubic lattice;
this geometry allow us to keep $f_c$ higher than the maximum volume fraction of {\em sc} and {\em bc} cubic lattice and in the same
time $f_c$ is below the maximum value $0.74$ which correspond to touching spherical cells.
Furthermore {\em fc } cubic lattice seems to be
the most realistic model for a living tissue \cite{Pavlin_fcc}. \\
}
%
\subsection{Effective Permittivity of a Single Cell}
\begin{figure}[h]
\centering
\includegraphics[scale=1.0, width=8cm]{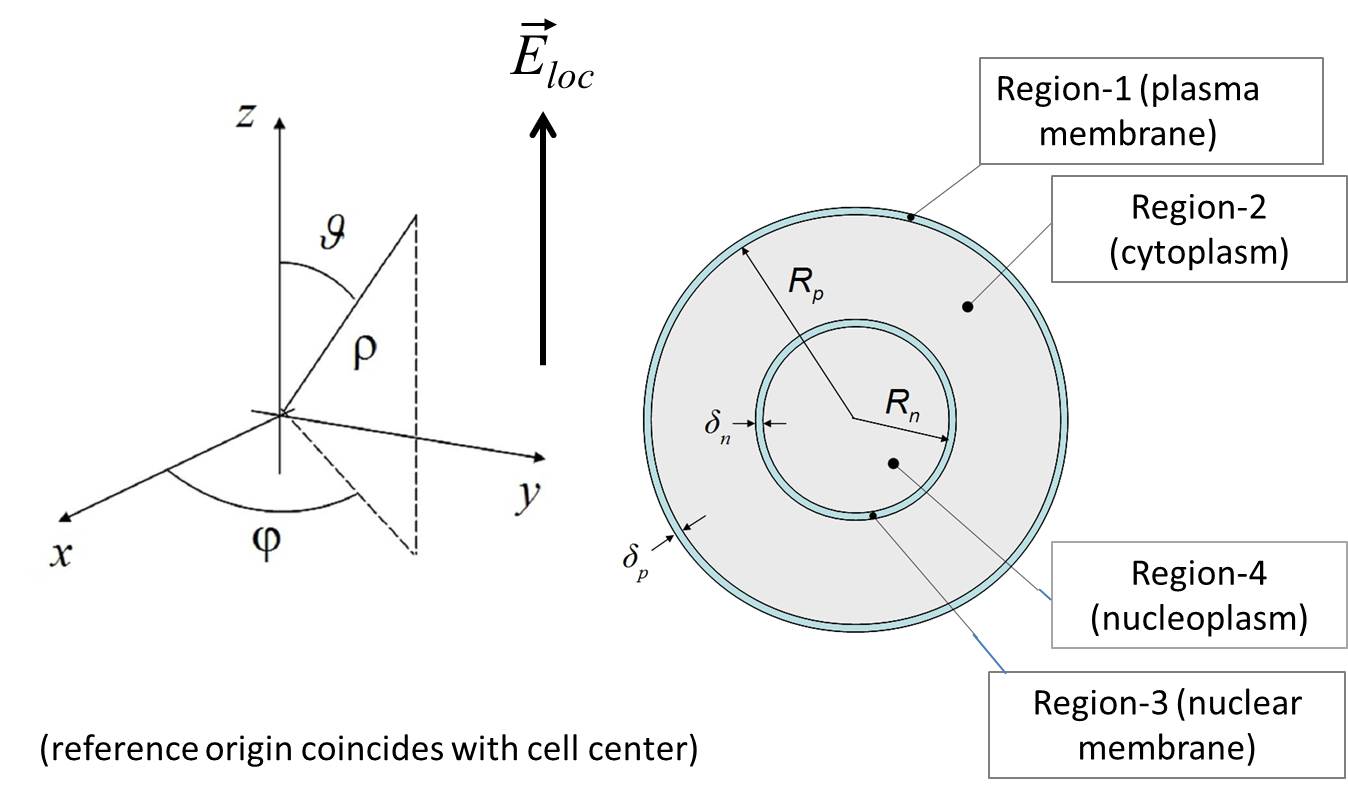}
\caption{Nucleated spherical cell ({\bfp four concentric layers}) exposed to a local field  $E_{loc}$. The origin of reference system is the cell center. } 
\label{fig6}
\end{figure}

%
A simple model of a living cell is  sketched in Figure \ref{fig6}: a stratified sphere, 
where $R_{p}$ is the radius of the cytoplasm, 
$\delta_{p}$  is the thickness of the plasma membrane, 
so that $R_c=R_p+\delta_p$,
$R_{n}$ is the radius of a spherical organelle (assumed concentric for simplicity), 
and $\delta_{n}$ is the thickness of the inner membrane.
%
%
In the spirit of EMT \cite{MixBook}, the multilayered cell depicted in Figure \ref{fig6}, 
can be modeled  as a {\em homogeneous} sphere with an equivalent permittivity $\epsilon_{c}$ . 
A simple EMT formula is available \cite{layered_sphere_Chen} for the effective permittivity 
of a homogeneous sphere coated by a {\it single} layer, 
\begin{equation}
\epsilon_{c}=\frac{1-2G}{1+G}\epsilon_{shell}
\label{eq:single_cell}
\end{equation}
where
\begin{equation}
G=\frac{\epsilon_{shell}-\epsilon_{core}}{2\epsilon_{shell}+\epsilon_{core}}\left(\frac{R_{core}}{R_{core}+\delta_{shell}}\right)^{3}
\end{equation}
where $\epsilon_{core}$,  and $\epsilon_{shell}$,  are the core and shell complex permittivities,
$R_{core}$  is  the core radius and  $\delta_{shell}$ the shell thickness.
This formula can be used {\it recursively}, by assuming the core to be in turn a {\it layered} sphere,
and $\epsilon_{core}$  its {\it effective} permittivity, to obtain an explicit formula for the
equivalent permittivity of the stratified cell in Figure \ref{fig6}, which is not reported for brevity.
%
\subsection{Effective Permittivity of Tissue}
%
Letting $\epsilon_h$ the permittivity of the host medium in which the cells are dispersed, 
we may use Bruggemann formula \cite{MixBook}
to derive the effective permittivity $\epsilon_{eff}$ of the bio-tissue  as follows
\begin{equation}
f_c \frac{(\epsilon_c-\epsilon_{eff})}{\epsilon_c+2\epsilon_{eff}}+
(1-f_c)\frac{(\epsilon_h-\epsilon_{eff})}{\epsilon_h+2\epsilon_{eff}}=0
\label{eq:eps_eff}
\end{equation}
where $\epsilon_c$ and $f_c$ are the equivalent cell permittivity and the cell volume fraction in the tissue, respectively.
We note in passing that Bruggemann formula, 
which treats the host and the inclusions in a symmetric way, 
is more appropriate for modeling a {\it densely} packed tissue, 
compared to the Maxwell-Garnett formula used in \cite{layered_sphere_Chen},
which is accurate only when $f_c \ll 1$ \cite{MixBook}.
%
\subsection{Average vs Local Field}
\begin{figure}[h]
\includegraphics[scale=1.0, width=8cm]{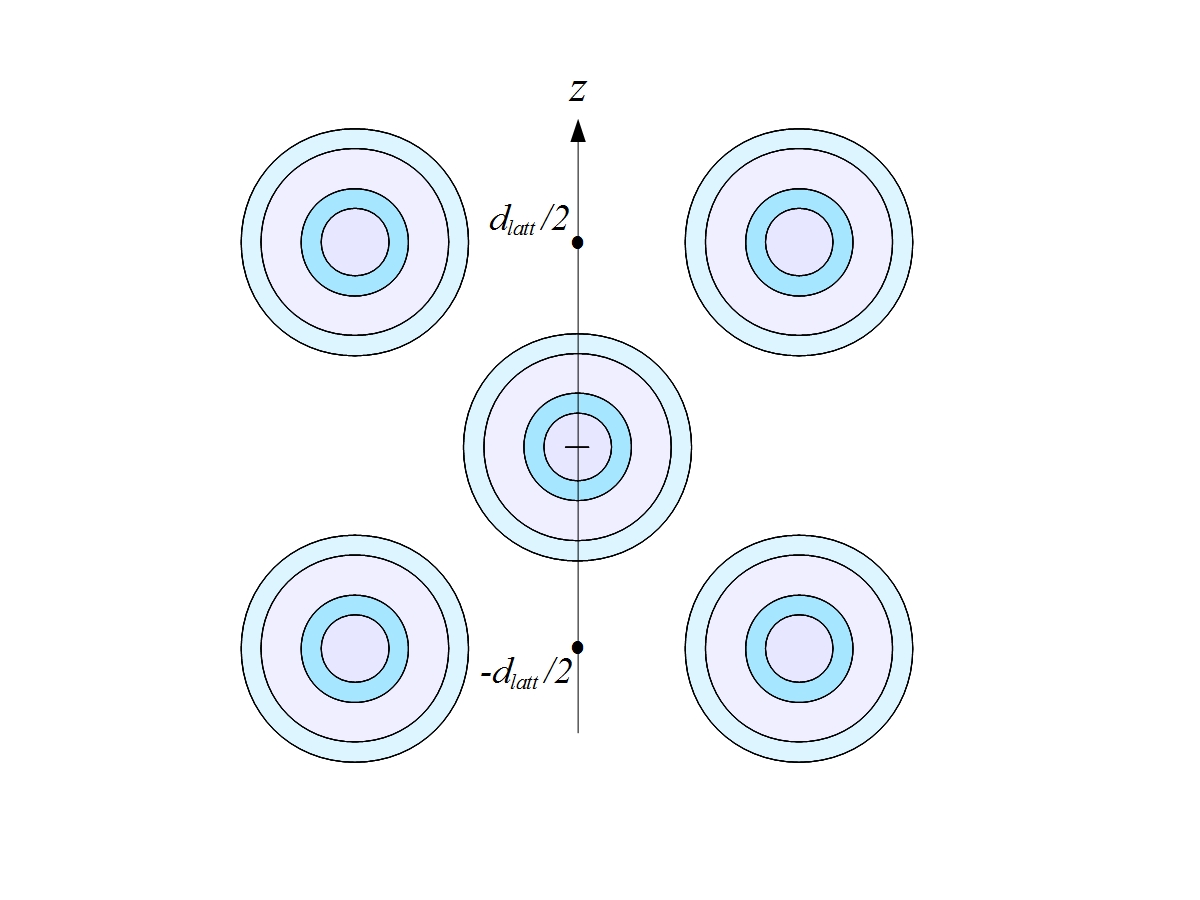}
\caption{{\bfp Separate cells } in {\it fc} cubic lattice model of tissue (side view), the lattice unit-cell side length 
$d_{latt}$ is displayed.  Relevant to equation (\ref{eq:pot_AB}).} 
\label{fig3}
\end{figure}

%
Let $E_{eff}$ the effective ({\it average}) field in the effective medium with permittivity $\epsilon_{eff}$ representing the tissue; 
$E_{loc}$  the  {\it local} field  to which a single cell is actually exposed (see Fig. \ref{fig6}),  
and  $E^{(e)}$ and $E^{(i)}$ the {\it local} external and intracellular fields including the cell response to $E_{loc}$. 
Under the quasistatic approximation, consistent with the use of EMT,
and assuming the effective and local fields as linearly polarized parallel to the lattice edge,
we may write following \cite{Qin} (see Fig. \ref{fig3} for the relevant notation)
\[
E_{eff} d_{latt}=
\int_{-d_{latt}/2}^{-R_c}~E^{(e)}_z(z)~dz
\]
\begin{equation}
+ \int_{-R_c}^{R_c}~E^{(i)}_z(z)~dz+
\int_{R_c}^{d_{latt}/2}~E^{(e)}_z(z)~dz\\
\label{eq:pot_AB}
\end{equation}
and \cite{JacksonBook}
\begin{equation}
\left\{
\begin{array}{l}
\displaystyle{
E^{(i)}_z(z)=\frac{3\epsilon_{h}}
{\epsilon_{c}+2\epsilon_{h}}
E_{loc}
},\\
\\
\displaystyle{
E^{(e)}_z(z)=\left[
\frac{2(\epsilon_{c}-\epsilon_{h})}
{\epsilon_{c}+2\epsilon_{h}}
\left(\frac{R_c}{z}\right)^{3}+1
\right]E_{loc}
},
\end{array}
\right.
\label{eq:in_out}.
\end{equation}
%
%
Equations (\ref{eq:pot_AB}), (\ref{eq:in_out}) can be combined 
to relate the effective and local fields as follows:
\begin{equation}
E_{loc}=\frac{
E_{eff}
}{
\displaystyle{
1+2\left(\frac{R_c}{d_{latt}}\right)
\left[\frac{\epsilon_{h}-\epsilon_{c}}{\epsilon_{c}+2\epsilon_{h}}\right]
}}.
\label{eq:E_local}
\end{equation}
{\bfp
It should be noted that the analysis carried out in this work holds only when the external 
electric fields are not strong or long enough to induce the formation of membrane pores.}

%
\subsection{Average vs Impressed Field}
%
Finally we consider a tissue specimen in vacuum. 
The effective field in the tissue $E_{eff}$  is related to the vacuum applied field $E_0$ in a way which depends
on the tissue specimen geometry and orientation with respect to the applied field.
We shall consider the case where the tissue  forms a  circular disc
(like a cultured cell layer in a Petri glass).  
Also, for the sake of simplicity, we shall assume that the disc diameter is also electrically small.
Thus, again in the quasi-static limit,
\begin{equation}
E_{eff}=\Xi(\epsilon_{eff},\epsilon_0,\epsilon_{sub})E_{0}
\label{eq:E_impressed}
\end{equation}
where $\epsilon_{sub}$ is the dielectric constant of the tissue substrate (e.g., glass).
The explicit form of the factor $\Xi(\epsilon_{eff},\epsilon_0,\epsilon_{sub})$  can be obtained 
in the quasi-static approximation using EMT, 
under suitable assumptions about the tissue and substrate dimensions.
For the simplest case where $\epsilon_{sub} \approx \epsilon_0$, 
$\Xi(\epsilon_{eff},\epsilon_0,\epsilon_{sub})=\epsilon_0/\epsilon_{eff}$ if the field is orthogonal to the tissue
and $\Xi(\epsilon_{eff},\epsilon_0,,\epsilon_{sub})=1$  if the field is parallel to the tissue.\\
Equations  (\ref{eq:E_local}) and (\ref{eq:E_impressed}) provide the needed relationship 
between the field $E_0$ applied to a tissue specimen,
and the local field $E_{loc}$ seen by each cell in the tissue.\\
All equations in this section can be rewritten in terms of the complex material conductivities,
modulo the usual substitution  $\epsilon \rightarrow \sigma/j\omega$.
%
\section{Fields and Powers in a Tissue Cell}
\label{sec:PowerDep.}
%
It is now possible to derive the electromagnetic field and dissipated power distribution 
for each cell embedded in a tissue, by solving the single (spherical, multilayer) cell boundary value problem 
whose geometry is sketched in Fig. \ref{fig6}.
In the quasi-static approximation, the total field can be derived from the (spectral) scalar potential 
\begin{equation}
\Phi(\vec{r},\omega)~=~\Phi_0(\vec{r})
-E_{loc}(\omega)~\rho \cos\vartheta~+~
\Phi_s(\vec{r},\omega),
\label{eq:fullpot}
\end{equation}
where $E_{loc}$ is the local field acting on the cell, 
$\Phi_0(\vec{r})$ is the resting potential
\begin{equation}
\Phi_0(\vec{r})~=~[1-U(\rho-R_n)]V_n+[1-U(\rho-R_p)]V_p, 
\label{eq:phi0}
\end{equation}
$U(\cdot)$ being Heaviside's step-function, 
$V_n$, $V_p$ the cytoplasmatic and nuclear transmembrane potentials, and
\begin{equation}
\frac{\Phi_s(\vec{r},\omega)}{E_{loc}}=
\left\{
\begin{array}{l}
\Xi(\omega)~\rho\cos\vartheta~~~~~~~~\rho~<~R_{n}\\
\Psi(\omega)~\rho^{-2}\cos\vartheta~~~~~\rho~>~R_{c}\\
\Theta(\omega)\rho^{-2}+\Sigma(\omega)\rho]\cos\vartheta~~~~R_{n}<\rho<R_{c}
\end{array}
\right.
\label{eq:phism}
\end{equation}
the (dipolar) cell-response potential.  
The functions $\Xi(\omega)$,~$\Psi(\omega)$,~$\Theta(\omega)$,
$\Sigma(\omega)$ are determined by enforcing continuity 
of the (radial, inward) current density 
across the cell and nuclear membranes, viz.
\begin{equation}
\tilde{\sigma}_{eff}(\omega)
\left.\partial_{\rho}\Phi\right|_{\rho=R_{p}+\delta_{p}}=
\tilde{\sigma_{2}}(\omega)
\left.\partial_{\rho}\Phi\right|_{\rho=R_{p}}=
\Upsilon_1(\omega)\delta\Phi_{1},
\label{eq:BC1}
\end{equation}
\begin{equation}
\tilde{\sigma_{2}}(\omega)
\left.\partial_{\rho}\Phi\right|_{\rho=R_{n}+\delta_{n}}=
\tilde{\sigma_{4}}(\omega)
\left.\partial_{\rho}\Phi\right|_{\rho=R_{n}}=
\Upsilon_{3}(\omega)\delta\Phi_{3},
\label{eq:BC2}
\end{equation}
where
\[
\delta\Phi_{1}(\vartheta,\omega)=
\Phi(R_{p}+\delta_{p},\vartheta,\omega)\!-\!\Phi(R_{p},\vartheta,\omega)-V_{p}=
\]
\begin{equation}
=\Phi_s(R_{p}+\delta_{p},\vartheta,\omega)\!-\!\Phi_s(R_{p},\vartheta,\omega),
\label{eq:BC3}
\end{equation}
and
\[
\delta\Phi_{3}(\vartheta,\omega)=
\Phi(R_{n}+\delta_{n},\vartheta,\omega)\!-\!\Phi(R_{n},\vartheta,\omega)-V_{n}=
\]
\begin{equation}
=\Phi_s(R_{n}+\delta_{n},\vartheta,\omega)\!-\!\Phi_s(R_{n},\vartheta,\omega),
\label{eq:BC4}
\end{equation}
are the transmembrane excess potentials of the plasma and nuclear membrane.
In equations (\ref{eq:BC1})-(\ref{eq:BC4}), 
$\tilde{\sigma}=\sigma-j\omega\epsilon$
is the complex frequency-dependent conductivity, and the
suffixes $eff$, $2$ and $4$ refer to the external (effective) medium, cytoplasm and nucleus, respectively.
$\Upsilon_1(\omega)$ $\Upsilon_{3}(\omega)$ are the cytoplasmatic and nuclear membrane specific admittances ($ohm^{-1}m^{-2}$). 
These latter can be written 
\begin{equation}
\Upsilon_1(\omega)=\tilde{G}_1(\omega)+j\omega\tilde{C}_1(\omega)=
\delta_p^{-1}\tilde{\sigma_1}(\omega)
\end{equation}
\begin{equation}
\Upsilon_3(\omega)=\tilde{G}_3(\omega)+j\omega\tilde{C}_3(\omega)=
\delta_n^{-1}\tilde{\sigma_3}(\omega)
\end{equation}
where $\tilde{G}$ and $\tilde{C}$ denote the pertinent membrane specific conductance and capacitance,
$\sigma$  the complex  membrane conductivity, 
and $\delta$ the membrane thickness. \\
Hereinafter the (carrier) frequency dependency of the spectral quantities is implicit and omitted
for lighter notation.
After some simple algebra, neglecting the membrana thicknesses, one gets
\begin{equation}
\Xi=
1-9\Upsilon_{1}\Upsilon_{3}R_{p}^{4}R_{n}\tilde{\sigma}_{eff} \Gamma^{-1}
\label{eq:xi}
\end{equation}
\begin{equation}
\Psi=
\frac{R_{p}^{3}}
{-2+6\Upsilon_{1}R_{p}\sigma_{2}\left[3\Upsilon_{3}R_{p}^{3}R_{n}+2(R_{p}^{3}-R_{n}^{3})\sigma_{2}\right]\Gamma }
\label{eq:psi}
\end{equation}
\begin{equation}
\Theta=-3\Upsilon_{1}R_{p}^{4}R_{n}^{3}\tilde{\sigma}_{2}\tilde{\sigma}_{eff} \Gamma^{-1}
\label{eq:teta}
\end{equation}
\[
\Sigma=\Bigl(3\Upsilon_{3}R_{p}^{3}R_{n}+2\tilde{\sigma}_{2}(R_{p}^{3}-R_{n}^{3})\Bigr) \cdot
\]
\begin{equation}
\Bigl(\Upsilon_{1}R_{p}(\tilde{\sigma}_{2}-\tilde{\sigma}_{eff})+2\tilde{\sigma}_{2}\tilde{\sigma}_{eff}\Bigr)
 \Gamma^{-1}
\label{eq:sigma}
\end{equation}
where we have defined the coefficient 
\[
\Gamma=
2\tilde{\sigma}_{2}\left[3\Upsilon_{3}R_{p}^{3}R_{n}+2(R_{p}^{3}-R_{n}^{3})\tilde{\sigma}_{2}\tilde{\sigma}_{eff}\right.
\]
\[
\left.
+\Upsilon_{1}R_{p}[3\Upsilon_{3}R_{p}^{3}R_{n}
(\tilde{\sigma}_{2}+2\tilde{\sigma}_{eff})
\right. 
\]
\begin{equation}
\left.
+2\tilde{\sigma}_{2}[R_{n}^{3}(\tilde{\sigma}_{eff}-\tilde{\sigma}_{2})+
R_{p}^{3}(\tilde{\sigma}_{2}+2\tilde{\sigma}_{eff})]]\right].
\label{eq:piercof}
\end{equation}
By using Eqs. (\ref{eq:xi})-(\ref{eq:sigma}) and $\Gamma$,  the potentials and fields are completely determined.\\
The power dissipated in the plasma membrane (region 1 in Figure \ref{fig6}) can now be derived as follows
\begin{equation}
\displaystyle{
P_{1}=
\int_{0}^{2\pi}\!d\varphi~
\int_{0}^{\pi}\tilde{P}_{1}(\vartheta)R_{p}^{2}
\sin\vartheta~d\vartheta,}
\label{eq:Pm}
\end{equation}
where
\begin{equation}
\displaystyle{
\tilde{P}_{1}(\vartheta)=\frac{1}{2}\mbox{Re}[\Upsilon_{1}]
\left|\delta\Phi_{1}\right(\vartheta)|^2
|E_{loc}|^2,
\label{eq:PmTilde}}
\end{equation}
yielding
\begin{equation}
\displaystyle{
\frac{P_{1}}{|E_{loc}|^2}=\frac{2\pi R_{p}^{2}}{3}~
\mbox{Re}[\Upsilon_{1}]
\left|\frac{\Psi-\Theta}{R_{p}^{2}}-\Sigma R_{p}\right|^2
\label{eq:powermem}.}
\end{equation}
Similarly, the power dissipated in the nuclear membrane (region 3 in Figure \ref{fig6}) is
\begin{equation}
\displaystyle{
\frac{P_{3}}{|E_{loc}|^2}=\frac{2\pi R_{n}^{2}}{3}~
\mbox{Re}[\Upsilon_{3}]
\left|\frac{\Theta}{R_{n}^{2}}+(\Sigma -\Xi)R_{n}\right|^2.
\label{eq:powerinnmem}}
\end{equation}
The power dissipated in the cytoplasm (region 2 in Figure \ref{fig6})  is 
$$
P_{2}=\frac{\mbox{Re}[\tilde{\sigma_{2}}]}{2}~
\int_{0}^{2\pi} d\varphi
\int_{0}^{\pi} \sin\vartheta d\vartheta
\int_{R_{n}}^{R_{p}}\left| \nabla\Phi(\vartheta) \right|^2
\rho^2 d\rho,
$$
yielding
$$
\frac{P_{2}}{|E_{loc}|^2}=\frac{2\pi \mbox{Re}[\tilde{\sigma_{2}}]}{3}
\left\{
\left(\frac{R_{p}^{3}-R_{n}^{3}}{3}
\right)
\left(
1+|\Sigma|^{2}-2\mbox{Re}[\Sigma^{*}]
\right)
\right.
$$
\begin{equation}
\left.
-\frac{4 |\Theta |^{2}}{3(R_{p}^{3}-R_{n}^{3})}
+4 \mbox{log} 
\left(\frac{R_{p}}{R_{n}}\right)
\mbox{Re}  \left\{\Theta [1-\Sigma ^{*}] \right\}
\right\}.
\label{eq:powercit}
\end{equation}
Finally, the power dissipated in the nucleoplasm (region 4 in Figure \ref{fig6}) is:
$$
P_{4}=\frac{\mbox{Re}[\tilde{\sigma_{4}}]}{2}~
\int_{0}^{2\pi} d\varphi
\int_{0}^{\pi}\sin\vartheta d\vartheta
\int_{0}^{R_{n}}
\left|\nabla\Phi(\vartheta)\right|^2
\rho^2 d\rho
$$
hence,
\begin{equation}
\frac{P_{4}}{|E_{loc}|^2}=\frac{2\pi}{3} R_{n}^{3}\mbox{Re}[\tilde{\sigma_{4}}]\left|1-\Xi \right|^{2}.
\label{eq:powerinncit}
\end{equation}
%
\subsection{Power Deposition in the presence of Membrane Dispersion}
\label{sec:PowerDmemD}
%
In the following we specialize eqs. (\ref{eq:powermem})-(\ref{eq:powerinncit}) to include membrane(s) specific capacitance
dispersion adopting the model \cite{DeVitaPS}, in two  cases:
a) both membranes are dispersive; b) only the nuclear membrane is dispersive.
The two models take into account the morphological differences
existing between the organelle membrane and the cell membrane \cite{Alberts}.
The model \cite{DeVitaPS}  introduces the functional form of 
$C_1(\omega)$ and/or $C_3(\omega)$ , for the positive frequency axis we have
\begin{equation}
C_i(\omega)=C_\infty - \frac{1}{2}(C_\infty-C_0) \mbox{Erfc}\left(\frac{\omega - \omega_0}{\Delta \omega}\right)
~~~~~i=1~\mbox{or}~3
\end{equation}
where $\mbox{Erfc}(\cdot)$  is the complementary error function.
Capacitance for negative frequencies are obtained by $C_i(-\omega)=C_i(\omega)$.
With reference to the case a), eqs. (\ref{eq:powermem}) to (\ref{eq:powerinncit})
reduce to:

\begin{equation}
\displaystyle{
\frac{P_{1}}{\left|E_{loc}\right|^2}=\frac{3}{2}\pi R_{p}^{4}~\mbox{Re}[\Upsilon_{1}],
\label{eq:powermemD}}
\end{equation}

\begin{equation}
\begin{array}{l}
\displaystyle{
~~~~~~~~~
\frac{P_{2}}{\left|E_{loc}\right|^2}=\frac{1}{2}\pi\tilde{\sigma}_{2} R_{p}^{5}\left|\frac{\Upsilon_{1}}{\tilde{\sigma}_{2}}\right|^{2},}
\label{eq:powercitD}
\end{array}
\end{equation}

\begin{equation}
\displaystyle{
\frac{P_{3}}{\left|E_{loc}\right|^2}=\frac{27}{8}\pi~R_{n}^{2}~
\mbox{Re}[\Upsilon_{3}]
\left|\frac{R_{c}R_{n}\Upsilon_{1}}{\tilde{\sigma}_{2}}\right|^2,
\label{eq:powerinnmemD}}
\end{equation}

\begin{equation}
\begin{array}{l}
\displaystyle{
\frac{P_{4}}{\left|E_{loc}\right|^2}=\frac{27}{8}\pi R_{n}^{3}\tilde{\sigma}_{4}\left|\frac{R_{c}R_{n}\Upsilon_{3}\Upsilon_{1}}{\tilde{\sigma}_{2}^{2}}\right|^{2}.}
\label{eq:powerinncitD}
\end{array}
\end{equation}

If only the inner membrane is dispersive as supposed in case b), 
eqs. (\ref{eq:powermem}) to (\ref{eq:powerinncit}) yield

\begin{equation}
\displaystyle{
\frac{P_{1}}{\left|E_{loc}\right|^2}=\frac{2}{3}\pi~R_{p}^{2}~
\mbox{Re}[\Upsilon_{1}]
\left|\frac{3\tilde{\sigma}_{eff}\tilde{\sigma}_{2}}{\Upsilon_{1}(2\tilde{\sigma}_{eff}+\tilde{\sigma}_{2})}\right|^2,
\label{eq:powermemND}}
\end{equation}

\begin{equation}
\begin{array}{l}
\displaystyle{
~~~~~~~~~
\frac{P_{2}}{\left|E_{loc}\right|^2}=2\pi\tilde{\sigma}_{2} R_{p}^{3}\left|\frac{\tilde{\sigma}_{eff}}{(2\tilde{\sigma}_{eff}+\tilde{\sigma}_{2})}\right|^{2},}
\label{eq:powercitND}
\end{array}
\end{equation}

\begin{equation}
\displaystyle{
\frac{P_{3}}{\left|E_{loc}\right|^2}=\frac{27}{2}\pi~R_{n}^{4}~
\mbox{Re}[\Upsilon_{3}]
\left|\frac{\tilde{\sigma}_{eff}}{(2\tilde{\sigma}_{eff}+\tilde{\sigma}_{2})}\right|^2,
\label{eq:powerinnmemND}}
\end{equation}

\begin{equation}
\begin{array}{l}
\displaystyle{
~~~~~~~~~
\frac{P_{4}}{\left|E_{loc}\right|^2}=\frac{3}{2}\pi R_{n}^{5}\tilde{\sigma}_{4}\left|\frac{9\Upsilon_{3}\tilde{\sigma}_{eff}}{2\tilde{\sigma}_{eff}+\tilde{\sigma}_{2}}\right|^{2}}.
\label{eq:powerinncitND}
\end{array}
\end{equation}
%
\section{Temperature Evolution}
We first study the thermal evolution of tissue under the adiabatic approximation 
i.e. the temperature field changes are slow
with respect to the EM field dynamics. We find that with suitable
EM carrier a relevant rapid increase in membrane temperature occurs while the
average cell temperature does not rise.
In the following section the full thermal responses analysis confirms the
order of magnitude calculation displaying quantitative features of proposed models.

\subsection{Adiabatic Approximation}
%
The power dissipated in each elementary cube of tissue in Figure \ref{fig1} is 
\begin{equation}
P_{eff}=\frac{\omega\mbox{Im}[\epsilon_{eff}]}{2}
|E_{eff}|^2 d_{latt}^3.
\end{equation}
Each elementary cube  contains  four cells, 
where the total dissipated power is 
\begin{equation}
P_{cells}=4(P_1\!+\!P_2\!+\!P_3\!+\!P_4).
\end{equation}
Accordingly, the fraction of power absorbed by the tissue 
which is actually dissipated in the cells  is  $\xi=P_{cells}/P_{eff}$.\\
The volume density $\dot{Q}_i$ of the power dissipated in each regions of the cell can be 
directly 
related to the {\it macroscopic} tissue Specific Absorption Rate (SAR) as follows:
\vspace*{-2mm}
\begin{equation}
\dot{Q}_i=\xi SAR \frac{
\displaystyle{
\eta_i+\sum_{j \neq i}^{1\dots 4} \frac{\eta_j V_j}{V_i}
}
}{
\displaystyle{
1+\sum_{j \neq i}^{1\dots 4} \frac{P_j}{P_i}
}}, \mbox{   }i={1\dots 4},
\label{eq:heatmem}
\end{equation}
where $\eta_i$ and $V_i$ are the pertinent mass densities $[kg/m^{3}]$ and volumes, respectively.\\
On time-scales much shorter than  the typical heat-diffusion
time constant $\tau_D$  across the cell subregions\footnote{
For an homogeneous material whose density, thermal conductivity, heat capacity
and thermal diffusion length are denoted as $\eta$, $\chi$, $c^{(p)}$ and $D$, 
$\tau_D=D^{2}\eta c^{(p)} /\chi$   \cite{Carslaw}.}
the  temperatures  $\Theta_i$  will evolve {\it adiabatically}, so that
\begin{equation}
\frac{\partial \Theta_{i}}{\partial t}
\approx
\frac{\dot{Q}_{i}}{\eta_{i}c^{(p)}_i},
\label{eq:adiabatically}
\end{equation}
where $c^{(p)}_i$ are the pertinent (constant-pressure) heat capacities.\\
Hence, for case a) in section \ref{sec:PowerDmemD}, we have
\begin{equation}
\frac{\partial \Theta_{1}}{\partial t}
\approx
\frac{\dot{Q}_{1}}{\eta_{1}c^{(p)}_{1}}
\approx
SAR\frac{R_{p}}{\delta_{p}}\frac{\eta_{2}}{\eta_{1}c^{(p)}_1}~\frac{\tilde{\sigma}_{2}~\mbox{Re}\!\!\ [\Upsilon_{1}]}{(3\tilde{\sigma}_{2}\mbox{Re}\!\!\ [\Upsilon_{1}]+R_{p}\left|\Upsilon_{1}\right|^{2})} 
\label{eq:appTm}
\end{equation}
\begin{equation}
\frac{\partial \Theta_{3}}{\partial t}
\approx
\frac{\dot{Q}_{3}}{\eta_{3}c^{(p)}_{3}}\approx
\frac{3}{4}SAR\frac{R_{p}R_{n}^{2}}{\delta_{n}}\frac{\eta_{2}}{\eta_{3}c^{(p)}_{3}}\frac{\mbox{Re}[\Upsilon_{3}]\left|\Upsilon_{1}\right|^{2}}{\mbox{Re}[\Upsilon_{1}]~\tilde{\sigma}_{2}^{2}}.
\label{eq:appTnm}
\end{equation}
\begin{equation}
\frac{\partial \Theta_{2}}{\partial t}
\approx
\frac{\dot{Q}_{2}}{\eta_{2}c^{(p)}_{2}}
\approx SAR\frac{R_{p}\left|\Upsilon_{1}\right|^{2}}{3c^{(p)}_{2}\mbox{Re}[\Upsilon_{1}]\tilde{\sigma}_{2}},
\label{eq:appTc}
\end{equation}
\begin{equation}
\frac{\partial \Theta_{4}}{\partial t}
\approx
\frac{\dot{Q}_{4}}{\eta_{4}c^{(p)}_{4}}
\approx\frac{9}{4}SAR\frac{\eta_{2}}{\eta_{4}c^{(p)}_{4}}\frac{R_{n}^{2}~R_{p}~\tilde{\sigma}_{4}\left|\Upsilon_{1}\right|^2\left|\Upsilon_{3}\right|^2}{\tilde{\sigma}_{2}^{4}~\mbox{Re}[\Upsilon_{1}]}.
\label{eq:appTnc}
\end{equation}
\\
Similarly, for case b) in section \ref{sec:PowerDmemD}, we have
\begin{equation}
\frac{\partial \Theta_{1}}{\partial t}
\approx
\frac{\dot{Q}_{1}}{\eta_{1}c^{(p)}_1}\approx
SAR\frac{\eta_{2}}{\eta_{1}c^{(p)}_1}
\frac{\tilde{\sigma}_{2}~\mbox{Re}[\Upsilon_{1}]}{\left|\Upsilon_{1}\right|^{2}},
\label{eq:appTmND}
\end{equation}
\begin{equation}
\frac{\partial \Theta_{2}}{\partial t}
\approx
\frac{\dot{Q}_{2}}{\eta_{2}c^{(p)}_{2}}
\approx SAR\frac{4\tilde{\sigma}_{2}R_{c}^{3}}{c^{(p)}_{2}(4\tilde{\sigma}_{2}R_{c}^{3}+27R_{n}^{4}~\mbox{Re}[\Upsilon_{3}])},
\label{eq:appTcND}
\end{equation}
\begin{equation}
\frac{\partial \Theta_{3}}{\partial t}
\approx
\frac{\dot{Q}_{3}}{\eta_{3}c^{(p)}_{3}}
\approx\frac{9}{4}SAR\frac{R_{n}^{2}}{\delta_{n}}\frac{\eta_{2}}{\eta_{3}c^{(p)}_{3}}\frac{\mbox{Re}[\Upsilon_{3}]}{\tilde{\sigma}_{2}},
\label{eq:appTnmND}
\end{equation}
\begin{equation}
\frac{\partial \Theta_{4}}{\partial t}
\approx
\frac{\dot{Q}_{4}}{\eta_{4}c^{(p)}_{4}}
\approx~SAR\frac{R_{n}^{2}}{c^{(p)}_{4}}\frac{\eta_{2}}{\eta_{4}}\frac{\tilde{\sigma}_{4}}{\tilde{\sigma}_{2}}\left|\Upsilon_{3}\right|^{2}.
\label{eq:appTncND}
\end{equation}
Some order of magnitude estimate can be obtained using (\ref{eq:appTm})-(\ref{eq:appTnc}) and  
(\ref{eq:appTmND})-(\ref{eq:appTncND}), together with the physical parameters  in Table \ref{table_A3} and \ref{table_A4}.
Furthermore, we assume $R_{p}=10^{-3}\mbox{m}$,  $R_{n}=3\cdot10^{-4}\mbox{m}$, 
$\delta_{p}=\delta_{n}=10^{-8}\mbox{m}$. \\
Results are very similar to formulas presented in \cite{DeVitaPS} and suggest
 the possibility that the inner and/or the outher membrane
temperature may rise up to lelvels at which biological damages  occurs.

\subsection{Thermal Response}
%
{\bfp 
In this subsection the tissue thermal response is obtained using the separate cells approximation valid for short (nanoseconds) EM pulsed field.
}
The time evolution of the temperature $\Theta$ in the cell is obtained by
solving the (coupled) heat diffusion equations
\begin{equation}
\displaystyle{
\eta_i c_i^{(p)}\frac{\partial \Theta_i}{\partial t}-\nabla\cdot\left(\chi_i\nabla \Theta_i\right)=\dot{Q}_i,\mbox{   }i=1,\dots,4
\label{eq:heateq}}
\end{equation}
$c_i^{(p)}$, $\chi_i$, $\dot{Q}_i$ being the heat capacity $[J/(kg^oK)]$, thermal conductivity $[W/(m^oK)]$ and EM-induced power density $[W/m^3]$.
For simplicity, we shall average\footnote{
Doing so, we neglect the possible relevance 
of {\it spatial} temperature gradients
in the membrane \cite{gradients}.}
the source term and the temperature distributions 
with respect to the angular variable $\vartheta$. 
The heat diffusion equation in the angle-averaged 
quantities becomes (using the same symbols $\Theta$, $\dot{Q}$
for the angle-averaged quantities, for notational ease)
\begin{equation}
\eta c_i^{(p)}\frac{\partial \Theta_i}{\partial t}-\frac{\chi_i}{\rho^{2}}\frac{\partial}{\partial \rho}
\left(\rho^{2}\frac{\partial \Theta}{\partial \rho}\right)=\dot{Q}_i.
\label{eq:tempsfe}
\end{equation} 
Equation (\ref{eq:tempsfe}) can be solved numerically,
starting from the initial conditions
\begin{equation}
\Theta_i(\rho,0)=\Theta_0, ~~~~~~\forall ~\rho
\end{equation} 
under the following boundary conditions: 
\begin{equation}
\left\{
\begin{array}{l}
\left.\Theta_i(\rho,t)\right|_{\rho=R_{i}}=
\left.\Theta_{i+1}(\rho,t)\right|_{\rho=R_{i}},
\\
\mbox{}
\\
\chi_i
\displaystyle{
\left.
\frac{\partial \Theta_i}{\partial \rho}
\right|_{\rho=R_{i}}
}=
\chi_{i+1}
\displaystyle{
\left.
\frac{\partial \Theta_{i+1}}{\partial \rho}
\right|_{\rho=R_{i}}
}
\end{array}
\right.,\mbox{   }i=1,\dots,4
\label{eq:contT}
\end{equation}
expressing the continuity of temperature and heat flux across the material boundaries
in Figure \ref{fig6}.
%
The domain where the solution is numerically sought is truncated 
using the further boundary condition 
\begin{equation}
\displaystyle{
\left.\frac{\partial \Theta}{\partial \rho}\right|_{\rho=R^*}=0,
\label{eq:adiab}}
\end{equation}
where $R^*$ is the radius of the cell-centered spheres in the tissue which touch without intersecting
(e.g., $R^*=2^{-3/2}d_{latt}$ for the {\em fc} cubic lattice.)
Equation (\ref{eq:adiab}) expresses the reasonable requirement that no heat flux exists 
between individual (neighbouring) cells in the tissue. 
{\bfp This is due to the negligible temperature diffusion on the typical EM pulses time scale.}
%
\section{Numerical Results.}
\label{num_results4}
%
In this section we illustrate some representative results 
based on numerical solution of the heat diffusion problem in Sect. III, 
obtained by using a finite element simulation code ($COMSOL^{\mbox{\textregistered}}$).
{\bfp 
The thermal response of a separate cell is azimuthally invariant. 
Due to the dipolar nature of EM excitation in the quasitatic approximation (on both inner and outer membranes) 
the thermal distribution takes the simple shape of $\cos^2 \vartheta$ (with reference to Fig. 2 for the angular variable). 
The membranes average values (that are the most relevant quantity)
 of these distributions are  displayed in Fig.s 4, 5, 6, 7.
}

\begin{figure}[h]
\centering
\includegraphics[scale=1.0, width=8cm]{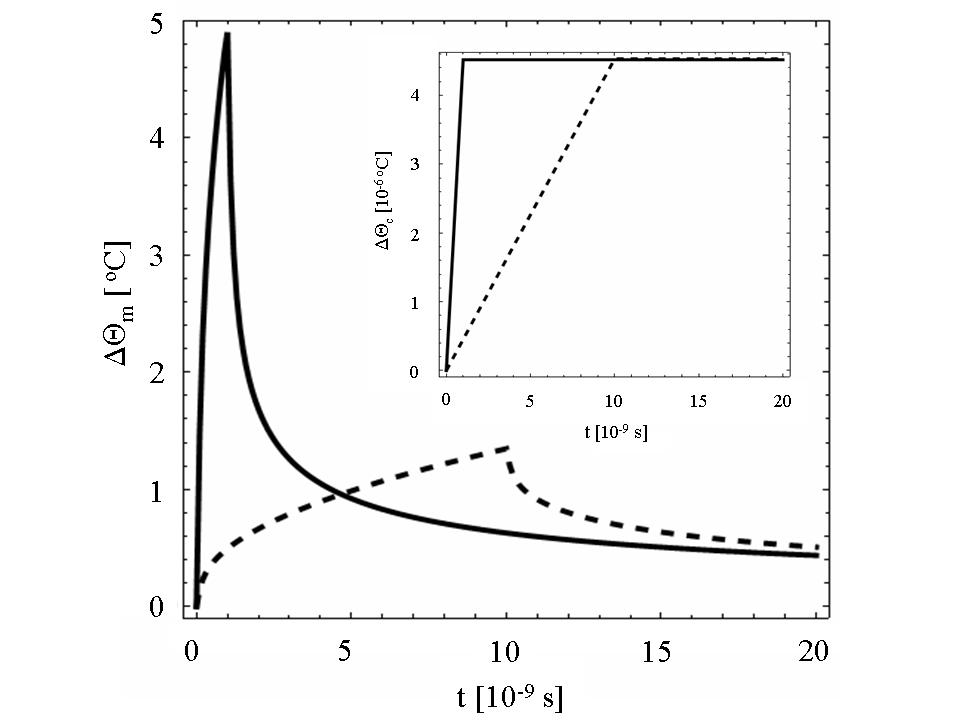}
\caption{Increase of the (average) {\em outer} membrane temperature of a spherical multilayer-cell 
model (case a) Section \ref{sec:PowerDmemD}) versus time, for an incident (rectangular) pulse with a Specific 
Absorption Dose of 1 ~J/kg, applied at t=0. (Full line) Pulse width = 1~ns. (Dashed line) 
Pulse width = 10~ns. Dispersion parameter for both membranes supposed equal are $C_0= 10^{-2}$ F/m$^2$, $C_\infty= 10^{-9}$ F/m$^2$,
$\omega_0 = 2\pi 10^8$ Hz, $\Delta \omega = 2 \pi 10^7$ Hz. {\bfp The cells volume fraction is $f_c=0.64$}. 
 The corresponding temperature increase in the cytoplasm is shown in the inset} 
\label{fig7}
\end{figure}

\begin{figure}[h]
\centering
\includegraphics[scale=1.0, width=8cm]{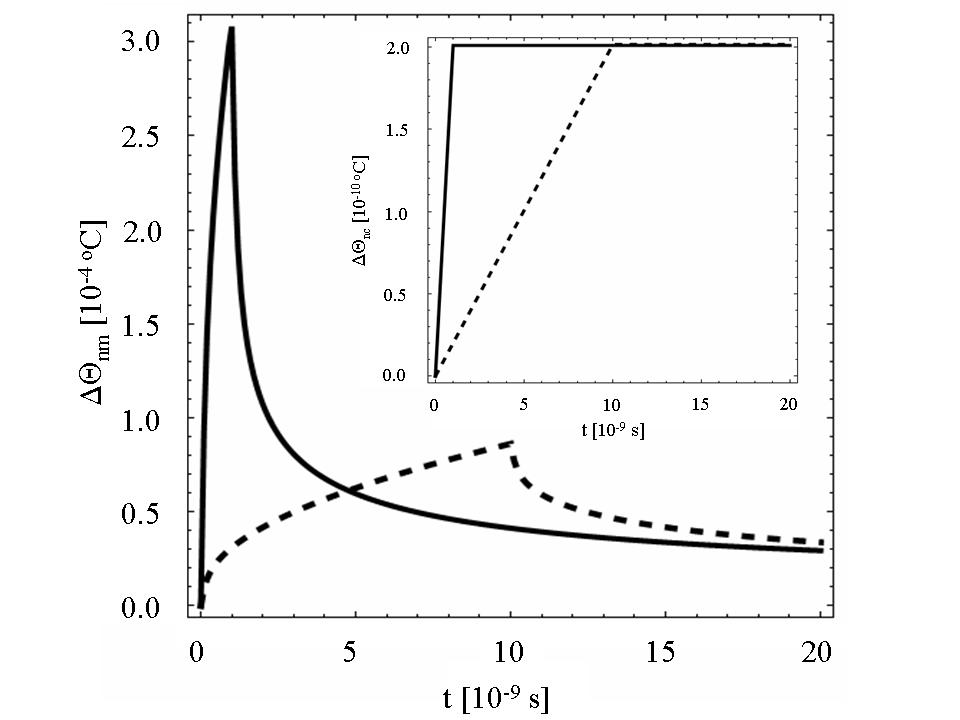}
\caption{Increase of the (average) {\em inner} membrane temperature of a spherical multilayer-cell model (case a) Section \ref{sec:PowerDmemD}) 
versus time, for an incident (rectangular) pulse with a Specific Absorption Dose of 1~J/kg, applied at t=0. (Full line) Pulsewidth = 1ns. 
(Dashed line) Pulsewidth = 10ns. The relevant dispersion parameter are reported in Figure \ref{fig7}.
{\bfp The cells volume fraction is $f_c=0.64$} 
The corresponding temperature increase in the nucleoplasm is shown in the inset} 
\label{fig8}
\end{figure}

\begin{figure}[h]
\centering
\includegraphics[scale=1.0, width=8cm]{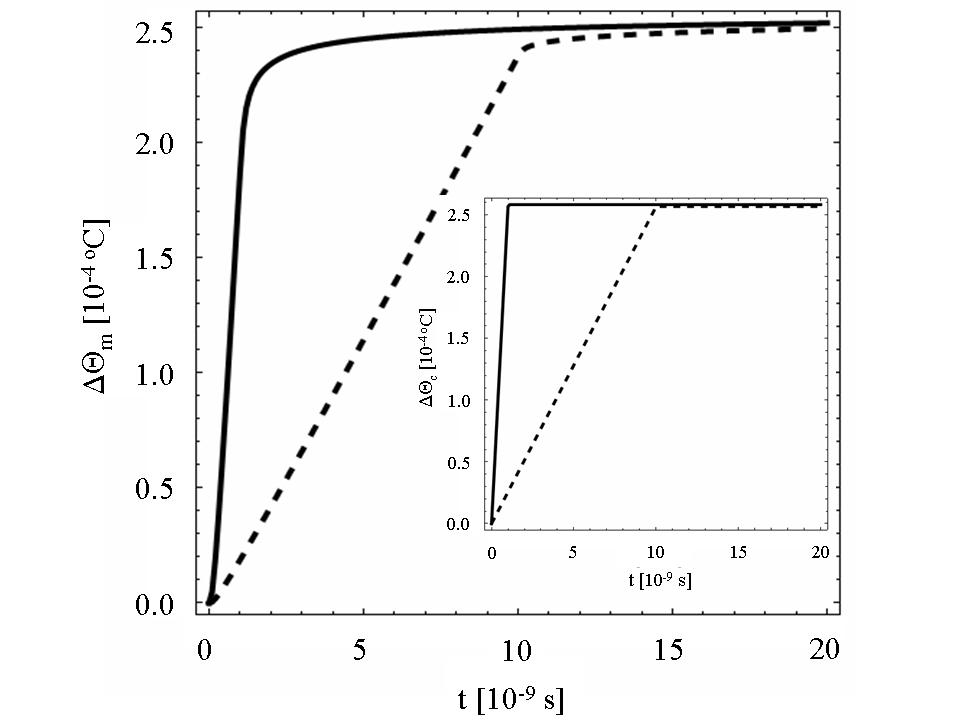}
\caption{Increase of the (average) {\em outer} membrane temperature of a spherical multilayer-cell 
model (case b) Section \ref{sec:PowerDmemD}) versus time, for an incident (rectangular) pulse with a Specific 
Absorption Dose of 1 J/kg, applied at t=0. (Full line) Pulsewidth = 1ns. (Dashed line) 
Pulsewidth = 10ns. This membrane is non dispersive.{\bfp The cells volume fraction is $f_c=0.64$}.
 The corresponding temperature increase in the cytoplasm is shown in the inset} 
\label{fig9}
\end{figure}

\begin{figure}[h]
\centering
\includegraphics[scale=1.0, width=8cm]{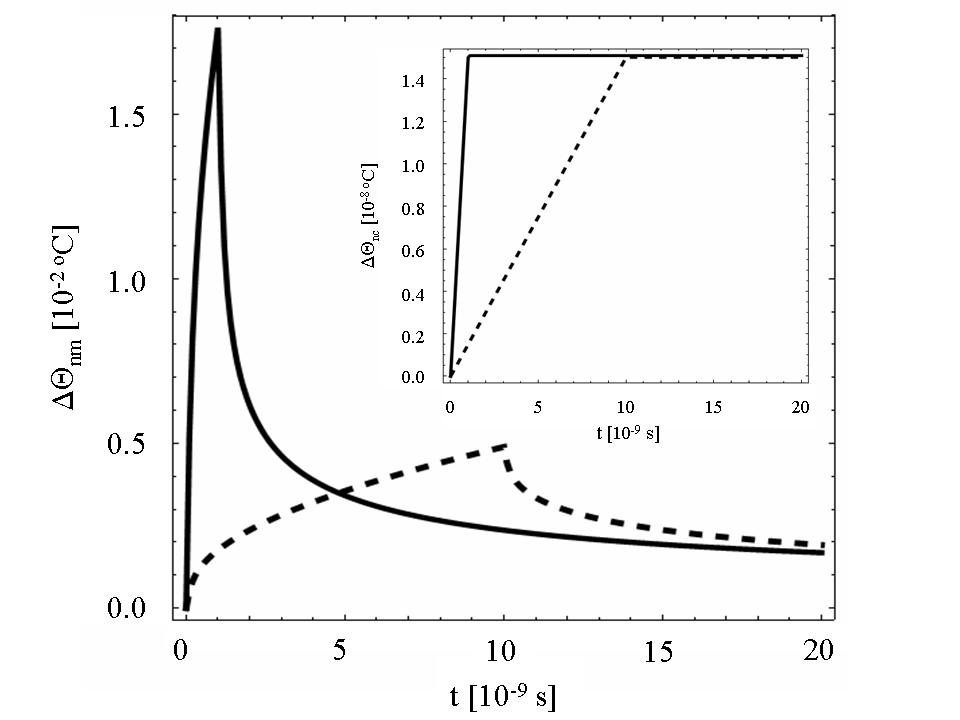}
\caption{Increase of the (average) {\em inner} membrane temperature of a spherical multilayer-cell 
model (case b) Section \ref{sec:PowerDmemD}) versus time, for an incident (rectangular) pulse with a Specific 
Absorption Dose of 1 J/kg, applied at t=0. (Full line) Pulsewidth = 1ns. (Dashed line) Pulsewidth = 10ns.
Dispersion parameter for nuclear membrane is $C_0= 10^{-2}$ F/m$^2$, $C_\infty= 10^{-9}$ F/m$^2$,
$\omega_0 = 2\pi 10^8$ Hz, $\Delta \omega = 2 \pi 10^7$ Hz. {\bfp The cells volume fraction is $f_c=0.64$}.
The corresponding temperature increase in the nucleoplasm is shown in the inset} 
\label{fig10}
\end{figure}

In our simulations we assume $R_{p}=10^{-3}\mbox{m}$,  $R_{n}=3\cdot10^{-4}\mbox{m}$, 
$\delta_{p}=\delta_{n}=10^{-8}\mbox{m}$. 
The electromagnetic and thermal parameters, 
from \cite{Kotnik} and \cite{gradients}, respectively, 
are reported in Tables \ref{table_A3} and \ref{table_A4}. 
The host extracellular medium is modeled as a  saline solution, 
using Stogryn  formulas \cite{Stogryn},  including fixes from \cite{Klein}. 
\renewcommand{\arraystretch}{2}
\begin{table}
\caption{Electromagnetic parameters of a multilayered living cell \cite{Kotnik}.} 
\label{table_A3}
\footnotesize
\begin{center}
\begin{tabular}{|c|c|c|c|}
\hline
$$&Conductivity &Permittivity\\
$$&[$S/m$]&[$As/Vm$]\\
\hline
${\mbox{Extracellular} \atop \mbox{medium}}$  &1.24 & $6.4 \cdot 10^{-10}$\\
\hline
${\mbox{Plasma and Nuclear} \atop \mbox{membrane}}$  & $3 \cdot 10^{-7}$ & $4.4 \cdot 10^{-11}$\\
\hline
${\mbox{Cytoplasm} \atop \mbox{Nucleoplasm}}$& 0.3 & $6.4 \cdot 10^{-10}$\\
\hline
\end{tabular}
\end{center}
\end{table}
%
\begin{table}
\caption{Thermal parameters of a multilayered living cell \cite{gradients}.} 
\label{table_A4}
\footnotesize
\begin{center}
\begin{tabular}{|c|c|c|c|}
\hline
$$&Heat capacity &Mass density~& ${\mbox{Thermal} \atop \mbox{~conductivity}}$\\
$$&[$J/(kg^oK)$]&[$kg/m^{3}$]&[$W/(m^oK)]$\\
\hline
${\mbox{Extracellular} \atop \mbox{medium}}$ &4 $ \cdot 10^{3}$&$10^{3}$&0.577\\
\hline
${\mbox{Plasma and Nuclear} \atop \mbox{membrane}}$ &2 $\cdot 10^{3}$ & 9$ \cdot 10^{2}$&0.2\\
\hline
${\mbox{Cytoplasm} \atop \mbox{Nucleoplasm}}$&4$ \cdot 10^{3}$&$10^{3}$&0.577\\
\hline
\end{tabular}
\end{center}
\end{table}
With reference to Section \ref{sec:PowerDmemD} case a), the temperature increase in the plasma cell membrane of a single multilayer 
cell model is shown in Fig. \ref{fig7}, for two different pulse widths (1ns and 10ns represented with a solid and dashed line, respectively) 
and a fixed Specific Absorption Dose (1 J/kg). 
As it can be seen, increasing the pulse duration at constant pulse energy produces a slower temperature raise.
The corresponding temperature raise in the cytoplasm (shown in the inset) is in both cases negligible compared to the membranes temperature. 
The temperature increase in the nuclear cell membrane is displayed in Fig. \ref{fig8} for two different pulse durations 
and in the inset the corresponding temperature increase in the nucleoplasm.
\noindent As it can be seen, the temperature increase in {\bfp the nuclear membrane is several
orders of magnitude lower than that in the plasma membrane, 
while the temperature increase in the nucleoplasm and cytoplasm is negligible compared to the membranes temperature}.
The same analysis performed in Figs. \ref{fig7} and \ref{fig8} is reported in Figs. \ref{fig9} and \ref{fig10} respectively, but specialized 
to the case b) discussed in Section \ref{sec:PowerDmemD}.
It is interesting to note that the temperature {\bfp increase in both the membranes is
well below the threshold where biological damage occurs, even if the nuclear membrane temperature increase
is about two orders of magnitude higher than that of the plasma membrane.
As a conclusive remark we note that the temperature increase in the nucleoplasm and cytoplasm are once again negligible with respect to the membranes heating.}

So, adopting a dispersive model for both the plasma and nuclear cell
plasma and nuclear cell membrane, a localized heating effect
is observed only on the plasma membrane while the nuclear
membrane remains essentially unaffected.
Such results suggest that although pulsed fields of nanosecond-subnanosecond duration interact mostly with the inner organelles, 
possible heating effects could be observed on the outer cell membrane.

Also in this more complicated but more realistic case, a steep relatively large 
temperature raise occurs in the {\bfp membranes}, without any significant temperature variation in the cytoplasm. 
\section{Conclusions}
\label{sec:concl}
We have presented a simple model of interaction between 
a bio tissue modeled as an assembly of multilayer cells and a pulsed EM field, 
by solving the coupled EM and thermal problem and including frequency 
dispersion in the nuclear and plasma membrane capacitances.
The complex permittivity of the tissue has been obtained using EMT via the classical Rayleigh formula, 
after computing the effective (complex) permittivity of each individual cell using the Lindell-Sihvola 
(closed form) EMT formula for stratified spheres. Once the effective permittivity of the tissue is known, 
we have computed the field in each spherical cell region in the quasi-static approximation, and then used the field
solution to deduce the time-dependent temperature distribution in all cell constituents, 
by solving numerically the related heat diffusion problem. 
{\bfp
The model considers densely but separated (no geometric/electric contact) cells in the tissue. 
The short duration (nanosecond) of the pulsed EM field results in a negligible spatial diffusion of the temperature field.
In this approximation, cells can be considered thermally independent and the condition (54),
ensuring absence of thermal flow, can be consistently used.
A more accurate thermal evolution, taking into account also transport phenomena (due to thermoregulation) ,
and the study of the cells geometric disposition effect on the thermal tissue dynamic, are out of 
the scope of this paper and will be faced in future works.
}
It is interesting to note {\bfp in this simplified model}
that by using short, large-amplitude pulses, whose spectral content overlaps significantly the membrane capacitance dispersion band,
and whose duration is shorter than the membrane thermal diffusion time, one may observe fast membrane localized heating 
(the cytoplasm and nucleoplasm temperatures being essentially
unaffected). Membrane heating may thus occur in the
absence of macroscopic heating of the whole cell (i.e., in a
{\em nonthermal} exposure regime) and may affect the membrane physiology.
The direct measurement of the membrane local temperature raise is, admittedly, a difficult issue.
One possibility could be to use a laser beam to probe the mechanical normal modes
of the membrane before and after pulsed excitations, to check the expected permanent
change in membrane stiffness due to membrane protein denaturation.
Whether the above findings may bear any relevance to explain,
even in part, any of the observed biophysical effects in living
cells exposed to subnanosecond MV/m pulsed fields remains to
be investigated and makes the case for strong interdisciplinary
cooperation.
\section{Appendix:~Formula derivation}
In this appendix we give further details on the derivation of (30)-(37).
This is accomplished for equations (33) and (37) starting from (28),
similar calculations can be applied to the other cases.
The (28) equation is
\[
\begin{array}{l}
{\displaystyle \frac{P_{4}}{|E_{loc}|^2}=\frac{2\pi}{3} R_{n}^{3}\mbox{Re}[\tilde{\sigma_{4}}]\left|1-\Xi \right|^{2}},
\end{array}
\]
where $\Xi$ is reported in (18) and the definition of the coefficient $\Gamma$ (contained in the exppression for $\Xi$) is given in (22).
These expressions can be simplified by noting that $R_n \ll R_p$, resulting
in the approximation $R_p^3 - R_n^3\sim R_p^3$.
In addition, the (complex) membrane specific admittances defined in (16) and (17)
 depending on the capacitance, drop to their lower values  in presence of strong membrane dispersion. 
Taking into account these assumptions, after a lenghtly but straightforward calculations we find
in the cases a)  (both membranes are dispersive): 
\[
\Gamma=4 \sigma_c^2 \sigma_{eff}R_c^3.
\]
The same computation gives in the case b) 
(only the nuclear membrane is dispersive) the result:
\[
\Gamma=4 \sigma_c R_c^4 \Upsilon_1(\sigma_c + 2\sigma_{eff}).
\]
The simplified equation of $\Gamma$ substituted in (28), after some simple
algebra , gives the equations (33) and (37) of the paper.

%


\begin{thebibliography}{99}




\bibitem{Garner}{V. L. Garner, M. Deminsky, V. B. Neculaes, V. Chashihin, A. Knizhnik and B. Potapkin, "Cell membrane thermal gradients induced by electromagnetic fields" {\em J. Appl. Phys.}, vol. 113, pp. 214701-1-214701-11, Jun. 2013.}


\bibitem{tutorial}{K. H. Schoenbach, R. P. Joshi, J. F. Kolb, N. Chen, M. Stacey, E. S. Buescher, S. J. Beebe and P. Blackmore, "Ultrashort electrical pulses open a new gateway into biological cells,"  {\em Proc. IEEE}, pp. 1122-1137, 2004.}

\bibitem{plasma_perm}{A. G. Pakhomov, J. F. Kolb, J. A. White, R. P. Joshi, S. Xiao and K. H. Schoenbach, "Long-lasting plasma membrane permeabilization in mammalian cells by nanosecond pulsed electric fields," {\em Bioelectromagnetics}, vol. 28, pp. 655-663, Jul. 2007.}

\bibitem{phosphat}{P. T. Vernier, Y. Sun, L. Marcu, C. M. Craft and M. A. Gundersen, "Nanoelectropulse induced phosphatidylserine translocation," {\em Biophys. J.}, vol. 86, pp. 4040-4048, Jun. 2004.}

\bibitem{zap}{K. H. Schoenbach, R. Nuccitelli and S. J. Beebe, "ZAP - extreme voltage could be a surprisingly delicate tool in the fight against cancer," {\em IEEE Spectrum}, pp. 20-26, Aug. 2006.}

\bibitem{melanoma}{R. Nuccitelli, U. Pliquett, X. Chen, W. Ford, R. J. Swanson, S. J. Beebe, J. F. Kolb and K. H. Schoenbach, "Nanosecond pulsed electric fields cause melanomas to self-destruct," {\em Biochem. and Biophys. Res. Commun.}, vol. 343, pp. 351-360, May 2006.}

\bibitem{other_cancers}{E. H. Hall, K. H. Schoenbach and S. J. Beebe, "Nanosecond pulsed electric fields induce apoptosis in p53-wildtype and p53-null HCT116 colon carcinoma cells," {\em Apoptosis}, vol. 12, pp. 1721-1731, May 2007.}

\bibitem{heating}{J. Van der Zee, "Heating the patient: a promising approach?," {\em Ann. Oncol.}, pp. 1173-1184, 2002.}

\bibitem{Schwan_1}{H. P. Schwan, "Electrical properties in tissue and cell suspensions," {\em Adv. Biol. Med. Phys.} vol. 5, pp. 147-209, 1957.}

\bibitem{Schwan_2}{H. P. Schwan and C. Grosse, "Alternating field evoked membrane potentials: effects of membrane and surface conductance," in {\em Proc. Annu. Int. Conf. EMBS} pp. 1523-1524, 1990.}

\bibitem{Schwan_3}{C. Grosse and H. P. Schwan, "Cellular membrane potentials induced by alternating fields," {\em Biophys. J.}, vol. 63, pp. 1632-1642, Dec. 1992.}

\bibitem{Haydon}{D. A. Haydon and B.W. Urban, "The admittance of the squid giant axon at radio frequencies and its relation to membrane structure," {\em J. Physiol.}, vol. 360, pp. 275-291, Mar. 1985.}

\bibitem{DeVitaPS}{R. P. Croce, A. De Vita, V. Pierro and I.M. Pinto, "A thermal model for pulsed EM field exposure effects in cells at non-thermal levels," {\em IEEE Trans. Plasma Sci.}, vol. PS 38, pp. 149-155, Feb. 2010.}

\bibitem{Song}{J. Song, R. P. Joshi and K. H. Schoenbach, "Synergistic effects of local temperature enhancements on cellular responses in the 
context of high-intensity, ultrashort electric pulses," {\em Med Biol Eng Comput.}, vol. 49, pp. 713-718, 2011.}

\bibitem{MixBook}{A. Sihvola, {\it Electromagnetic Mixing Formulas and Application}, IEE Press, London, 1999.}

\bibitem{tissue_cubic}{M. Pavlin, T. Slivnik and D. Miklav$\check{c}$i$\check{c}$, "Effective conductivity of cell suspensions," {\em IEEE Trans. Biomed. Eng.}, vol. BME 49, pp. 77-80, Jan. 2002.}

\bibitem{Pavlin_fcc}{M. Pavlin and D. Miklav$\check{c}$i$\check{c}$, "The effective conductivity and the induced transmembrane potential in dense cell system exposed to DC and AC electric fields," {\em IEEE Trans. Plasma Sci.}, vol. PS 37, pp. 99-106, Jan. 2009.}

\bibitem{layered_sphere_Chen}{L. F. Chen, C. K. Ong and T. G. Tan, "Effective permittivity of layered dielectric sphere composites," {\em J. of Materials Science}, vol. 33, pp. 5891-5894, Dec. 1998.}

\bibitem{Qin}{Y. Qin, S. Lai, Y. Jiang, T. Yang and J. Wang, "Transmembrane voltage induced on a cell membrane in suspensions exposed to an alternating field: A theoretical analysis," {\em Bioelectrochemistry}, vol. 67, pp. 57-65, Febr. 2005.}

\bibitem{JacksonBook}{J.D. Jackson, {\it Classical Electrodynamics}, J. Wiley \& Sons, New York, 1975, Sect. 4.4.}

\bibitem{Alberts}{B. Alberts, A. Johnson, J. Lewis, M. Raff, K. Roberts, P. Walter, 
{\em Molecular Biology of the Cell.} Oxford, Taylor \& Francis Inc. 2007.}


\bibitem{Carslaw}{H. S. Carslaw and J. C. Jaeger, {\em Conduction of Heat in Solids.} Oxford University Press, NY, USA, 2001.}



\bibitem{gradients}{A. W. Friend, S. L. Gartner, K. R. Foster and H. Howe, "The effects of high power microwave pulses on red blood cells and the relationship to transmembrane thermal gradients," {\em IEEE Trans. Microwave Theory Tech.} vol. MTT 29, pp. 1271-1277, Dec. 1981.}

\bibitem{Kotnik}{T. Kotnik and D. Miklavcic, "Theoretical evaluation of voltage inducement on internal membranes of biological cells 
exposed to electric fields," {\em Biophys. J.}, vol. 90, pp. 480-491, Jan. 2006.}


\bibitem{Stogryn}{A. Stogryn, "Equations for calculating the dielectric constant of saline water," {\em IEEE Trans. Microwave Theory Tech.} vol. MTT 19, pp. 733-736, Aug. 1971.}
\bibitem{Klein}{L.A. Klein and C.T. Swift, "An improved model for the dielectric constant of sea water at microwave frequencies," {\em IEEE Trans. Antennas Propagat.}, vol. AP 25, pp. 104-111, Jan. 1977.}
\end{thebibliography}
\end{document}